Christopher K. Allsup

Aurametrix, USA

https://aurametrix.com

# Stretching Rubber, Not Budgets: Accurate Parking Utilization on a Shoestring

## Abstract

Effective parking management is essential for ensuring safety and convenience in master-planned communities, particularly in active adult neighborhoods experiencing rapid growth. Accurately assessing parking utilization is a crucial first step in planning for future demand, but data collection methods can be costly and labor-intensive. This paper presents a low-cost yet highly accurate methodology for measuring parking utilization using pneumatic road tubes connected to portable traffic counters from JAMAR Technologies, Inc. By integrating results from JAMAR's analysis tool with custom Python scripting, the methodology enables precise parking lot counts through automated parameter optimization and error correction. The system's efficiency allows for scalable deployment without significant manual observation, reducing both costs and disruptions to daily operations. Using Tellico Village as a case study, this effort demonstrates that community planners can obtain actionable parking insights on a limited budget, empowering them to make informed decisions about capacity expansion and facility scheduling.

## 1. Introduction

Adequate parking is a cornerstone of master-planned communities, particularly in active adult lifestyle neighborhoods experiencing significant population growth. Sufficient parking ensures safety and accessibility for residents and visitors, while also supporting local businesses. A proactive approach to parking management begins with measuring lot utilization to gain insights into both existing and future needs. Predicting peak usage at buildout [1] enables community and facility managers to take proactive measures, such as expanding lot capacity or adjusting facility schedules, before parking issues arise. By addressing potential issues early, planners can ensure seamless operations and sustained resident satisfaction.

## 2. Tellico Village Parking Study

The case study for this work focuses on Tellico Village, an active adult lifestyle community with more than 5,800 homes located in East Tennessee. As the population of Tellico Village continued to grow, residents and guests began encountering parking congestion at various facilities. At the request of the Property Owners Association (POA) Board of Directors, the Long-Range Planning Advisory Committee (LRPAC) conducted a study to assess the scope of the problem.

In April 2024, the LRPAC conducted a brief survey to identify which facilities were most affected. Responses from over 1,100 residents revealed that the average percentage of time residents reported having "difficulty finding a parking space" was highest at the Wellness Center (14%), the Toqua clubhouse (19%), Chota Recreation Center (28%), and the Yacht Club (29%). Although the survey was not designed to uncover the full extent of parking issues, the data indicated that 12–14% of respondents experienced parking difficulties at least three times per month.

Based on these findings, the decision was made to collect more detailed data to determine whether parking capacity at the four facilities was sufficient to meet current and future demand. Specifically, utilization measurements of each parking lot would enable the LRPAC to:

a) Assess the severity, timing, and duration of insufficient parking capacity, and
b) Estimate capacity requirements through buildout.

The estimate for (b) is agnostic, as overutilization can be addressed by either increasing capacity (adding parking spaces to meet demand), reducing peak demand (incentivizing visitors to arrive at other times), or a combination of both approaches.

## 3. Seasonality of Demand

Previous studies of visitor attendance at these facilities indicated that certain months experienced the highest average demand. Consequently, parking lot measurements were conducted only during those months. Instead of attempting to predict the specific days with the highest congestion, parking demand for each lot was measured continuously over one or more months, with the measurement period divided into multiple counts, each typically lasting approximately two weeks.

Once all measurements for a lot were recorded, utilization metrics were calculated as described in Section 13. These metrics were then used to estimate coverage levels, defined as the proportion of days or shorter intervals during which parking demand remains within available capacity, for both current conditions and buildout. Buffer requirements were also calculated and represent the number of additional parking spaces, or cars shifted outside peak periods, needed to meet a coverage target (see the Appendix).

## 4. Measurement Method

The LRPAC researched various vehicle counting methods that utilized different sensor technologies. Based on the project's goals, timeframe, and resources, the options were narrowed down using the following selection criteria: ease of deployment and data processing; ability to collect data continuously (24/7) over extended periods and in all weather conditions; capability to distinguish between cars and golf carts; ability to monitor multiple lots simultaneously; extendibility to other use cases; and cost-effectiveness for purchase and maintenance. Ultimately, portable traffic counters utilizing road tubes were selected as the measurement method, as they were the only option that satisfied all the criteria.

The Tellico Village POA purchased a traffic recording system designed and manufactured by JAMAR Technologies, Inc., based in Hatfield, PA. The system included two portable TRAX Pinnacle traffic counters [2], a single-use license for STARnext [3] (JAMAR's analysis software for processing traffic data), and road tubes with fixtures for installation.

## 5. Detection, Analysis and Post-Processing

Figure 1 illustrates the essential elements of the measurement process, which are explained below.

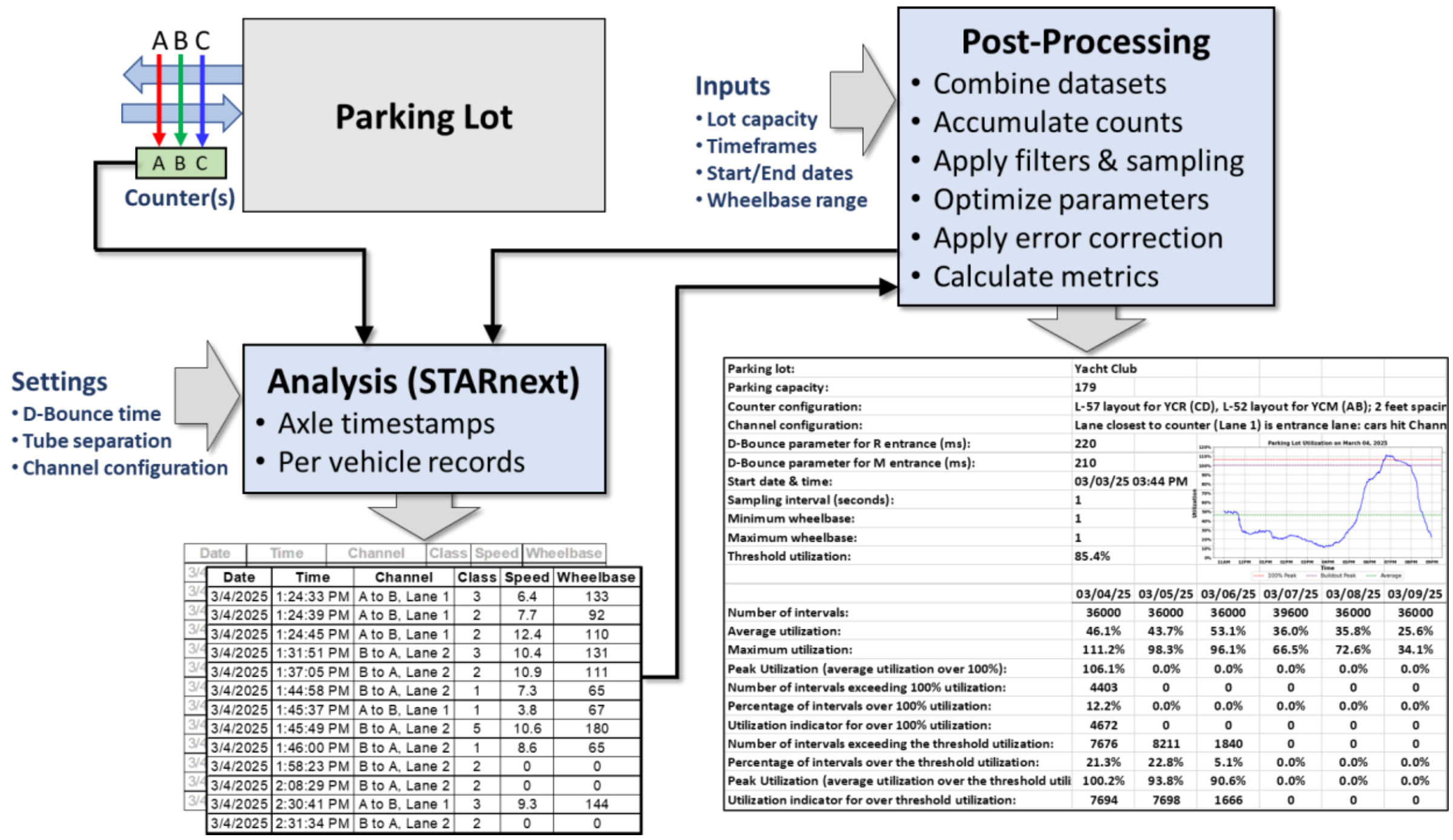

**Figure 1.** The measurement of parking utilization involved three major steps: detection, analysis, and post-processing.

**Detection.** When a vehicle's tires pass over a set of tubes, air pressure waves travel through the tubes to the traffic counter. The counter senses these pulses, converts them into electrical signals, and makes a timestamp recording for every axle that passes over the road tubes. This information is stored in memory along with related data for counting and subsequent analysis. Two counters are required when there are two separate entrances to a parking lot.

**Analysis.** Once the data from a count is transferred from the traffic counter to the STARnext analysis software, computations are performed based on user-controlled parameter settings. The primary objective of this step is to produce a set of axle timestamps that ensures the most accurate vehicle count. After this is achieved, a "golden" Excel worksheet is generated for the count. The worksheet contains columns for each detected vehicle, including the date, time, direction of travel (entry into or exit from the lot), and other per-vehicle details such as vehicle class, estimated speed, and wheelbase length. In cases where two counters are used, an analysis is performed for each count.

**Post-Processing.** This step, implemented in Python, processes the per-vehicle traffic data from the Excel worksheet(s) to produce an accurate utilization estimate for each day of the count. It combines datasets from multiple counters (if applicable), accumulates the counts, applies filters and sampling, optimizes parameters, applies error correction, and calculates various utilization metrics. An Excel worksheet is generated with separate sheets for each count, each including daily summary statistics. Additionally, plots are created that depict each day's utilization patterns.

## 6. Tube Configuration

For the most accurate results, the STARnext software requires data from a minimum of two tubes to estimate a vehicle's speed and wheelbase length. These estimations are based on the distance between the tubes, typically 2 feet, and the timing of the activation pulses they generate. The counter has four channels with input ports labeled A, B, C, and D that can be connected to up to four tubes. By convention, the lane closest to the counter is designated as Lane 1. Vehicles traveling in Lane 1 pass over the A tube (connected to the A channel) first, followed by the B tube (A → B). Vehicles in the opposite lane, Lane 2, approach the B tube first (B → A).

Distinguishing between vehicles entering and exiting a parking lot is essential. To maintain consistency across all measurements, we configured STARnext to assign Lane 1 for vehicles entering a lot and Lane 2 for vehicles exiting, as illustrated in Figure 2.

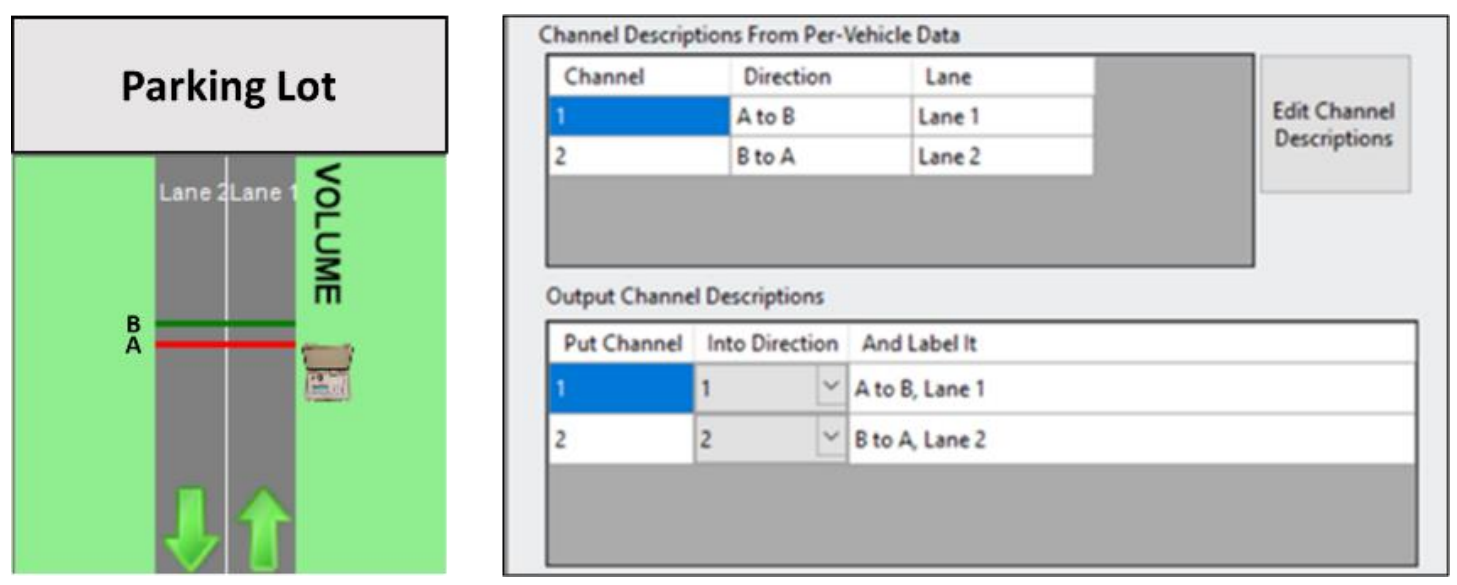

**Figure 2.** *Left*: The STARnext sensor layout showing two tubes connected to the A and B channels. Vehicles entering the parking lot use Lane 1. *Right*: Channel descriptions ensure correct directionality during post-processing by assigning the A → B direction to Lane 1 and the B → A direction to Lane 2.

A third tube (C) was added to provide fault tolerance in case one of the tubes becomes damaged during a count. STARnext supports the creation of custom sensor layouts to process data from any two channels independently. For example, if the A tube fails, a layout configuration that uses the B and C channels can be specified, allowing STARnext to calculate timestamp data using those channels.

If the B tube is damaged early in a count, which happened on several occasions at the Wellness Center parking lot, the count can still be salvaged by reconfiguring the sensor layout in STARnext to process data from the A and C channels. In this special case, the tube separation setting in STARnext was increased to 4 feet to reflect the physical distance between the undamaged A and C tubes.

Efforts should be made to eliminate counting errors that could compromise the accuracy of a count. Following best practices when setting up road tubes, including maintaining uniform spacing, proper tension, and equal tube lengths, helps minimize the number of unclassified vehicles in the dataset. In addition to these considerations, we applied innovative error minimization and correction techniques to achieve highly accurate estimates of parking lot demand, even in the presence of systemic detection errors. The following sections detail the methods employed to achieve this.

## 7. Predicted Parking Demand

Once the channel configuration and tube separation settings are specified in STARnext, we fine-tune the vehicle count by comparing the predicted parking demand with the observed parking demand at the end of the count. Before describing this fine-tuning process, we will first explain how parking demand is calculated during the post-processing step.

We partition a count into $N$ sampling intervals of equal duration $T$, expressed in seconds or minutes, and determine the number of timestamps recorded for each direction in each interval $i$. *Predicted parking demand* $D(n)$ represents the estimated number of vehicles in a parking lot after $n$ sampling intervals. Alternatively, $D$ can be described as a function of time $t$, where $t$ is an integer multiple $n$ of the sampling interval: $t = nT$. The value $D(t)$ is calculated as the initial observed number of vehicles, $D(t = 0)$, plus the cumulative difference between the number of vehicles entering and exiting the lot over all intervals up to $n$:

$$(1) \quad D(t) = D(nT) = D(t = 0) + \sum_{i=1}^{n} [\text{enter}(iT) - \text{exit}(iT)] \quad n \in 1, 2, \dots, N$$

Here, $\text{enter}(i)$ and $\text{exit}(i)$ are the numbers of vehicles entering or exiting the lot during sampling interval $i$. The precision of $D(t)$ depends on the duration of the sampling interval, with the minimum being $T = 1$ second, equivalent to the timing resolution of the counter.

Sampling intervals of 5 minutes or more are useful for computing rolling buffer requirements or simulating the effects of demand smoothing. For instance, one-hour fitness classes that allocate a fixed amount of time at the beginning of each hour for attendees to arrive and depart help mitigate large spikes in parking demand by spreading usage over the interval.

Predicted demand for a lot with two entrances is calculated as the sum of the cumulative differences between vehicles entering and exiting at each entrance. In practice, the calculation is applied using Equation (1) after combining the timestamp data from both counters into a single, chronologically ordered data frame.

## 8. Fine-Tuning Process

The “D-Bounce time” defines the duration the air switch in the TRAX Pinnacle counter waits after detecting a pulse before processing another. If the D-Bounce setting is suboptimal, extraneous pulses detected by the counter may be misinterpreted. These pulses can result from various phenomena. For instance, vehicles entering and exiting a parking lot may cross the tube system simultaneously, causing signal cross-talk that generates false positives. Additionally, a slow-moving vehicle crossing a tube at an angle may cause the left and right tires to pass over the tube fractions of a second apart, producing extra pulses. By adjusting the D-Bounce setting in STARnext, users can override the setting applied during the count to improve the accuracy of the recorded data.

The fine-tuning process for achieving highly accurate parking demand estimates begins with selecting a D-Bounce time $\delta$ that minimizes the percentage of unclassified vehicles in STARnext. This process continues through an automated, closed-loop interaction between STARnext and a custom Python script, as depicted in Figure 1. The script selects the D-Bounce time that minimizes the cumulative error: the difference between the predicted and observed number of vehicles in the lot at the end of a count.

In addition to adjusting the D-Bounce setting, vehicles with wheelbase lengths within a specified range $w$ can be excluded from a count to further reduce error. STARnext assigns a wheelbase length of zero if it cannot reliably classify a vehicle, so it is important not to filter out these unclassified vehicles unintentionally.

The iterative process leading to the optimal values $\delta^*$ and $w^*$ was fully automated using pywinauto [4] and PyAutoGUI [5], which are Python modules that allow users to control the mouse and keyboard programmatically through scripts. The parameter optimization is symbolically represented as:

$$(2) \quad (\delta^*, w^*) = \arg\min_{(\delta,w)} |D(N) - \Delta(N)|$$

where $N$ again represents the total number of sampling intervals in the count, and $D(N)$ and $\Delta(N)$ denote the predicted and observed values of the parking demand at the last interval, respectively.

## 9. Error Correction

Error correction is applied to each day of a count to improve the accuracy of the demand predictions. The calculation removes two error components from the predicted demand $D(t)$, whose cumulative error was minimized in the parameter optimization step described above. In the following explanations, we assume that a count begins on Day 0, and utilization is calculated for Days 1 through $M$.

The first error component is the cumulative error present at the start of Day j, which is equivalent to the sum of the accumulated errors incurred each previous day $\varepsilon_i$. The second error component is the error that accumulates over Day j itself, $\varepsilon_j$. *Adjusted predicted demand* for this day $D'_j(t)$ is:

$$(3) \quad D'_j(t) = D_j(t) - \left[\sum_{i=0}^{j-1} \varepsilon_i + \varepsilon_j\right] = D_j(t) - \sigma_j \quad j \in 1, 2, \dots, M$$

where $\sigma_j$ represents the cumulative error at the end of Day j. This error is the difference between the *predicted* end-of-day demand from calculations ($D_j$) and the *expected* end-of-day demand based on assumptions ($ED_j$):

$$(4) \quad \sigma_j = D_j - ED_j$$

The cumulative error at the end of the last day of the count is:

$$(5) \quad \sigma_M = D_M - \Delta_M$$

where $\Delta_M$ represents the ground truth observation of the number of vehicles in the lot at the end of the count.

The *accumulated* error incurred on the ith day is the difference between the *cumulative* errors:

$$(6) \quad \varepsilon_i = \sigma_i - \sigma_{i-1} = (D_i - ED_i) - (D_{i-1} - ED_{i-1})$$

and $\varepsilon_0 = D_0 - ED_0$ is the error at the end of Day 0. Substituting the expression for the accumulated error in Equation (6) into Equation (3) cancels out all the accumulated error terms up through Day j – 1, leaving only the cumulative error at the end of Day j.

Error correction applies a “phantom” demand vector, subtracting the cumulative error for each day $\sigma_j$ in the count ($j = 1, 2, \dots, M$) from the predicted demand $D_j(t)$. Figure 3 illustrates the predicted demand for a single day, both before and after error correction.

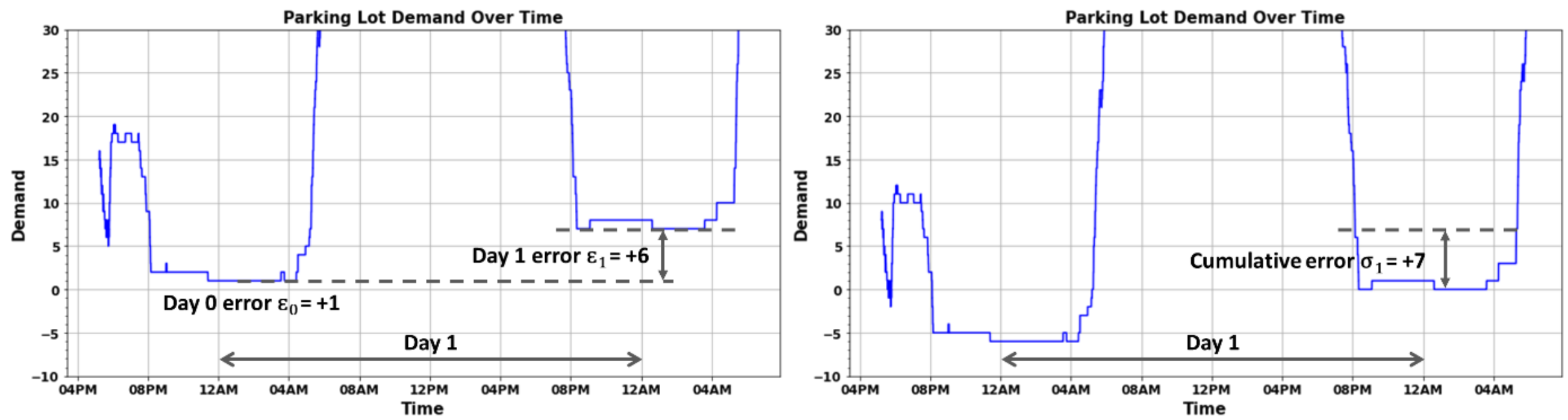

**Figure 3.** *Left*: The accumulated error on Day 0 is 1 vehicle ($\varepsilon_0$, ending at 1 a.m. on Day 1), and 6 on Day 1 ($\varepsilon_1$, ending 24 hours later), assuming $ED_0 = ED_1 = 0$. The cumulative error at the end of Day 1 is the sum of the accumulated errors, equivalent to $D_1 - ED_1 = 7$ vehicles. *Right*: Day 1's cumulative error is subtracted from its predicted end-of-day demand. The observed uniform decrease in demand on Day 0 can be disregarded, as the error correction only applies to hours of peak demand in Day 1.

For parking lots at facilities with daytime hours, the end-of-day time can be chosen to be late in the evening or early in the morning of the following day when the expected demand is zero. In this case, the error correction for each day simplifies to $-D_j$. However, an inspection of traffic datasets from the lots in our study revealed that they were not always empty at "dead-of-night," as shown in Figure 4. Moreover, it is possible that one or more vehicles could have remained parked for days without being detected.

| Date | Time | Channel |
|---|---|---|
| 1/14/2025 | 8:15:24 PM | C to B, Lane 2 |
| 1/14/2025 | 8:15:53 PM | C to B, Lane 2 |
| 1/14/2025 | 8:16:01 PM | C to B, Lane 2 |
| 1/14/2025 | 8:16:12 PM | C to B, Lane 2 |
| 1/15/2025 | 12:11:23 AM | B to C, Lane 1 |
| 1/15/2025 | 3:31:25 AM | B to C, Lane 1 |
| 1/15/2025 | 3:46:09 AM | C to B, Lane 2 |

**Figure 4.** Although the Wellness Center closed at 8 p.m. on January 14, two vehicles entered the parking lot later at 12:11 a.m. and 3:31.a.m.

Even though determining the ground truth demand $\Delta$ may not be possible, one can revise expected demand upward from zero as needed to improve accuracy. For example, if we choose 3:45 a.m. as the dead-of-night time for estimating end-of-day demand, the sequence of axle timestamps in Figure 4 suggests that at least two vehicles were in the lot at that time ($\Delta \geq 2$), making $ED = 2$ a more informed estimate of the expected demand. Observing dead-of-night activity in the dataset reduces false negatives compared to simply assuming that the expected delay is zero; moreover, false negatives occur only when contributions to the ground truth are unobservable in the timestamp data. Creating a Python script to analyze timestamps in this manner was straightforward, enabling the automatic generation of phantom demand vectors for error correction.

## 10. Avoiding Manual Counts

Discussions in previous sections have implied that manually counting vehicles in a parking lot before and after a traffic count is necessary for an accurate demand prediction. In fact, manual counting can be entirely avoided by estimating demand at all the dead-of-night times, as described in the previous section. Utilization results with and without manual counting are virtually identical, with the only notable difference being that the latter method excludes one day from each count.

During the fine-tuning process, the number of vehicles in the parking lot at the start of a count can be initialized to zero. Additionally, Equation (5) is augmented to account for the fact that the cumulative error at the end of the final day of the count (Day $M - 1$) is now based on the expected dead-of-night demand, rather than the ground truth observation of the number of vehicles in the lot the following day (Day $M$):

$$(7) \quad \sigma_{M-1} = D_{M-1} - ED_{M-1}$$

Two months into the study, we adopted this augmented approach for measuring all parking lot counts.

## 11. Least Squares Parameter Optimization

The cumulative cost function in Equation (2) represents a simple sum of the daily accumulated errors. We improved the accuracy of our estimates by replacing this with a sum-of-squares cost function, leveraging all observation points (i.e., the dead-of-night times) to select values for the D-Bounce time $\delta$ and wheelbase length range $w$ that minimize the sum of *squared* accumulated errors $\varepsilon_j$:

$$(8) \quad (\delta^*, w^*) = \arg\min_{(\delta, w)} \sum_{j=1}^{M-1} \varepsilon_j{}^2$$

Least squares parameter optimization prevents error cancellation and places greater emphasis on large errors, which can be advantageous in certain situations. For example, excluding golf carts from the Toqua parking lot counts, as discussed in Section 15, required increasing the upper limit of the wheelbase length beyond the typical range. The steep gradient of the sum-of-squares cost function imposed a reasonable constraint on w*, resulting in a more accurate demand calculation than that produced by minimizing only the cumulative error at the end of a count.

## 12. Peak Demand Error

There may be systemic sources of error that compromise prediction accuracy, even after applying error correction. For example, at the Wellness Center parking lot, vehicles slowly entering and exiting simultaneously during peak demand periods generated substantial signal cross-talk that had to be interpreted by a single counter. Despite these noisy conditions, the error correction effectively compensated for the high error rates and produced peak demand predictions that remained within the margin of error of manual counts—provided that the tubes were spaced apart by 2 feet as recommended for low-speed traffic.

However, recurring faults in the B tube over a period of at least four weeks required the use of channels A and C for the corresponding counts. The increased 4-foot separation between the tubes significantly raised the number of unidentified pulses, resulting in a predicted peak utilization that was approximately 7% higher than the manually observed value.

To avoid over- or underestimating peak demand in the presence of systemic data errors, we directly observe demand at or near its maximum on the highest-demand day of the week during an active traffic count. The *peak demand error* is defined as the difference between the adjusted predicted demand $D'$ from Equation (3) and the observed demand $D_{obs}$ at the peak time $t_p$. If the observation is made on Day $k$ of the count, then:

$$\text{(9)} \quad \text{Peak demand error} = D'_k(t_p) - D_{obs}(t_p)$$

The enhanced error correction applies a phantom demand vector that subtracts both this peak error and the cumulative error for each day j in the count from the predicted demand $D_j(t)$. Table 1 displays data from the peak error measurements of each parking lot. Uncertainties arising from manual counts of demand are indicated, along with the corresponding ranges within which the manual observations of peak utilization $U_{obs}$ were considered valid.

| Parking Lot | Lot Capacity | Observed Demand ($D_{obs}$) | Observed Utilization ($U_{obs}$) | Predicted Demand (D') | Predicted Utilization (U') | Tube Separation | Peak Demand Error |
|---|---|---|---|---|---|---|---|
| Wellness Center | 232 | 183 ± 3 | 78.9% ± 1.3% | 199 | 85.8% | 4′ | 16 |
| | | | | 181 | 78.0% | 2′ | -2 |
| Toqua | 170 | 143 ± 2 | 84.1% ± 1.2% | 143 | 84.1% | 2′ | 0 |
| Chota | 89 | 84 ± 1 | 94.4% ± 1.1% | 82 | 92.1% | 2′ | -2 |
| Yacht Club | 179 | 173 ± 2 | 96.6% ± 1.1% | 173 | 96.6% | 2′ | 0 |

**Table 1.** Peak error measurements for each parking lot. For the Wellness Center, demand was adjusted by -16 for counts with 4 feet of tube separation, and by +2 for counts with 2 feet.

Because our primary interest was in estimates from high-demand days, measuring peak error on just one such day per lot was sufficient, provided all tubes were functioning properly during that count. This condition allowed us to apply two distinct phantom demand adjustments, depending on whether a particular count used 2 or 4 feet of tube separation.

## 13. Utilization Metrics

In this section, we present the key utilization metrics employed in the Tellico Village parking study.

**Utilization** is defined as the ratio of parking lot demand to parking lot capacity:

$$\text{(10)} \quad U(t) = \frac{D'(t)}{C}$$

where $U(t)$ represents utilization at time t, $D'(t)$ is the error-corrected predicted lot demand, and C is the lot capacity.

**Maximum utilization** can serve as a useful initial indicator. If it regularly exceeds 100%, even for a short time, it may suggest a parking problem. In smaller parking lots, visitors do not take much time to search for spaces, so estimates of maximum utilization may not significantly exceed lot capacity.

**Average utilization**, measured over the course of open business hours, is another important metric. Predictably high average utilization is a positive sign for a business and can serve as a proxy for attendance. For instance, a restaurant could use historical average parking utilization data, combined with factors like weather, menus, and promotions, to forecast guest turnout and adjust resources accordingly.

The **percentage of time that utilization exceeds a specific threshold utilization $U_0$** offers additional insights into potential parking issues. For example, if demand exceeds $U_0 = 97\%$ of a lot's capacity for 30% of a 10-hour day, it may indicate a problem with parking availability.

**Peak utilization** $U_{peak}$ is defined as average utilization that exceeds a specified threshold utilization $U_0$:

$$(11) \quad U_{peak} = \frac{1}{\tau}\sum_{i=1}^{\tau}[U(t) > U_0]_i$$

where $\tau$ is the number of samples during which utilization exceeds $U_0$. This metric is particularly useful as it can help estimate the number of additional parking spaces needed or, equivalently, the average excess demand that must be shifted to less congested times to avoid adding spaces:

$$(12) \quad \text{Average excess demand} = D_{peak} - C = C \cdot (U_{peak} - 1)$$

where $D_{peak}$ is the peak demand.

If there are sampling intervals during which $U(t) > 100\%$, peak utilization is calculated using $U_0 = 100\%$ in Equation (11). For example, if a lot has 100 spaces and $U_{peak} = 120\%$, then the average excess demand of 20 vehicles suggests that the lot would need 20 additional spaces to meet peak demand. Alternatively, redistributing the 20 vehicles to non-peak times could be a more cost-effective solution.

If there are no sampling intervals during which $U(t) > 100\%$, peak utilization is calculated using $U_0 \leq 100\%$ in Equation (11). For this study, we selected $U_0$ to be the ratio of current households to platted lots. This ensures that if $U_{peak}$ is initially $U_0$, the average excess demand is expected to remain below zero until buildout, at which point $U_{peak}$ reaches 100% and excess demand becomes zero.

The **peak utilization indicator** $\Omega$ facilitates relative comparisons across peak utilization measurements by weighting the peak utilization $U_{peak}$ by the number of samples $\tau$ in which utilization exceeds the threshold utilization $U_0$:

$$(13) \quad \Omega = U_{peak} \cdot \tau$$

For example, if Monday's peak utilization is 120% based on $\tau = 120$ samples, and Tuesday's peak utilization is 140% based on $\tau = 80$ samples, Monday would represent the worst-case utilization because its $\Omega$ value is larger (144 versus 112).

## 14. Utilization Analysis: Yacht Club Parking Lot

The March 2025 utilization measurements for the Yacht Club parking lot were organized into four separate counts. Figures 5 and 6 present the results from the first count.

| | 03/04/25 | 03/05/25 | 03/06/25 | 03/07/25 | 03/08/25 | 03/09/25 |
|---|---|---|---|---|---|---|
| Number of intervals: | 36000 | 36000 | 36000 | 39600 | 36000 | 36000 |
| Average utilization: | 46.1% | 43.7% | 53.1% | 36.0% | 35.8% | 25.6% |
| Maximum utilization: | 111.2% | 98.3% | 96.1% | 66.5% | 72.6% | 34.1% |
| Time of maximum utilization: | 6:48 PM | 12:43 PM | 5:59 PM | 12:19 PM | 2:19 PM | 12:03 PM |
| Peak Utilization (average utilization over 100%): | 106.1% | 0.0% | 0.0% | 0.0% | 0.0% | 0.0% |
| Number of intervals exceeding 100% utilization: | 4403 | 0 | 0 | 0 | 0 | 0 |
| Percentage of intervals over 100% utilization: | 12.2% | 0.0% | 0.0% | 0.0% | 0.0% | 0.0% |
| Utilization indicator for over 100% utilization: | 4672 | 0 | 0 | 0 | 0 | 0 |
| Number of intervals exceeding the threshold utilization: | 7676 | 8211 | 1840 | 0 | 0 | 0 |
| Percentage of intervals over the threshold utilization: | 21.3% | 22.8% | 5.1% | 0.0% | 0.0% | 0.0% |
| Peak Utilization (average utilization over the threshold utilization): | 100.2% | 93.8% | 90.6% | 0.0% | 0.0% | 0.0% |
| Utilization indicator for over threshold utilization: | 7694 | 7698 | 1666 | 0 | 0 | 0 |
| | | | | | | |
| **Average Excess Demand Calculations** | | | | | | |
| Peak utilization: | 106.1% | 93.8% | 90.6% | 0.0% | 0.0% | 0.0% |
| Excess demand: | 11 | 0 | 0 | 0 | 0 | 0 |
| Peak utilization at buildout: | 124.2% | 109.8% | 106.0% | 0.0% | 0.0% | 0.0% |
| Excess demand at buildout: | 43 | 18 | 11 | 0 | 0 | 0 |

**Figure 5.** Yacht Club parking lot utilization data for the week beginning March 4, 2025.

Maximum utilization occurred on Tuesday, March 4 at 6:48 p.m., with $U_{peak} = 106.1\%$ and an average excess demand of 179 spaces × 6.1% = 11 cars. As of March 2025, Tellico Village had reached 85.4% of its planned buildout. This suggests that by the time the community reaches full buildout, $U_{peak}$ could reach 106.1% ÷ 85.4% = 124.2%, corresponding to an average excess demand of 43 cars. These vehicles may need to be redirected to less congested days or times; otherwise, an equivalent number of additional parking spaces would be required to accommodate the anticipated increase in congestion. It is also worth noting that utilization exceeded 100% for 12.2% of the hours the facility was open on Tuesday, equivalent to approximately 73 minutes.

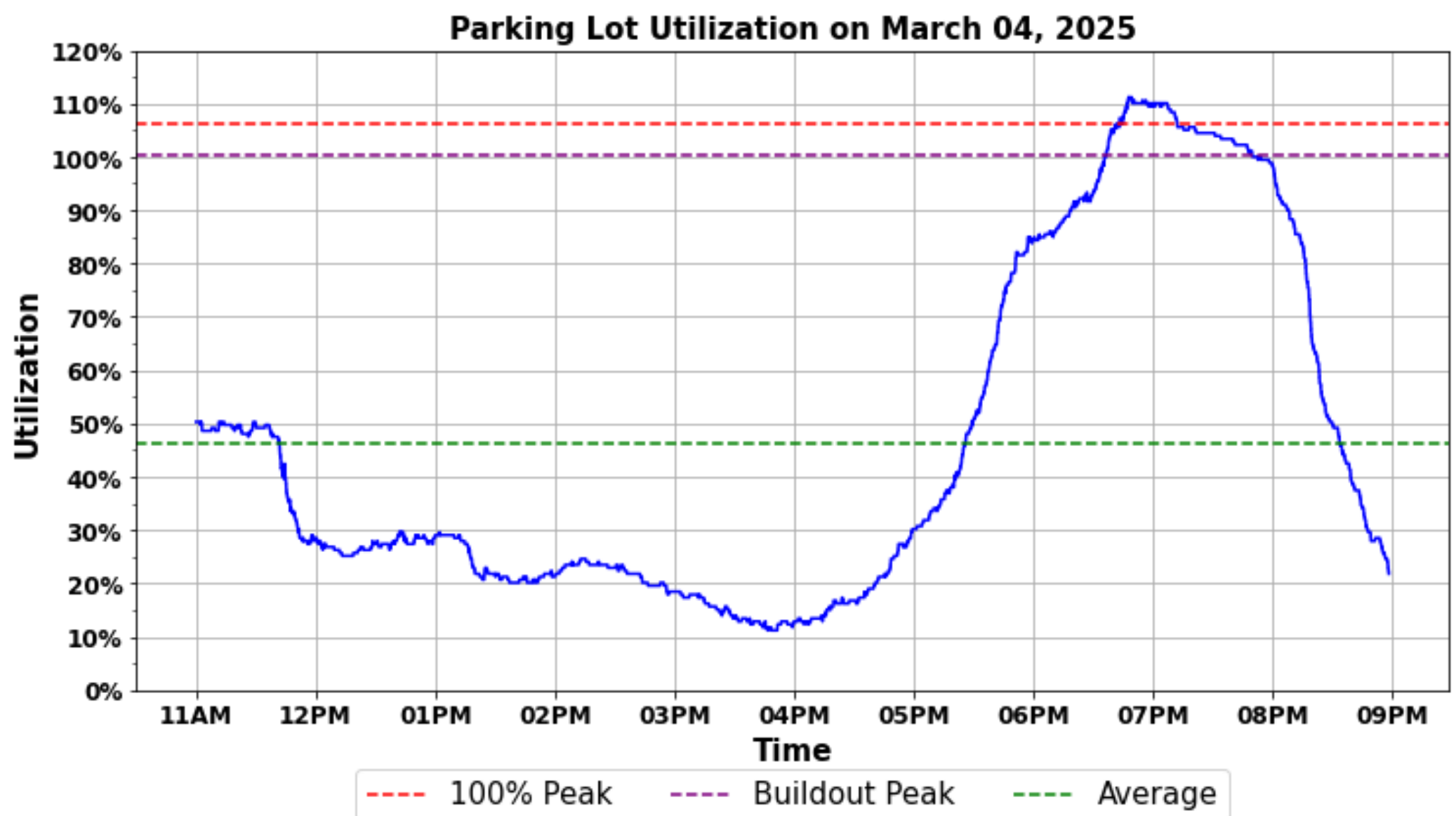

**Figure 6.** Parking utilization during Yacht Club business hours on March 4, 2025. Horizontal dotted lines indicate average utilization exceeding $U_0 = 100\%$ (red), average utilization exceeding $U_0 = 85.4\%$ (purple), and average utilization (green). Since maximum utilization exceeded 100%, only $U_0 = 100\%$ applies this day.

Maximum utilization did not exceed 100% from March 5 through March 9, indicating that no additional spaces are *currently* needed to meet parking demand on those days. Using a threshold utilization of $U_0 = 85.4\%$, peak utilization was highest on Wednesday, March 5, at 93.8%. This suggests that at full buildout, peak utilization could reach 109.8%, corresponding to an average excess demand of 18 cars.

## 15. Parking Lots

Each parking lot presented specific challenges and corresponding solutions, which are discussed below.

### Wellness Center

With 232 spaces, the Wellness Center parking lot is the largest in Tellico Village. The facility experiences some of its busiest days in January, as visitors aim to get back into shape after the holidays. Measuring the lot during winter weather posed unique challenges.

For instance, we had to temporarily remove the tube system in anticipation of a heavy snowstorm and subsequent snow plowing. Additionally, we discovered that mastic tape applied to the tubes becomes ineffective at temperatures below 30°F, as even light precipitation seeps under the strips, loosening the adhesive bond. Finally, traffic entering and exiting the parking lot during open hours, along with early darkness in winter after closing, made setting up the road tubes and starting a count potentially unsafe.

To address these issues, we developed a method for installing and removing the tubes in under 10 minutes. First, each galvanized clamp is nailed to the pavement, but the loop is left uncompressed to allow the tube to slide through easily, as shown in Figure 7. Next, a plastic zip tie is tightened around one end of each tube. When it is time to start the count, the tubes are threaded through the clamps and secured at the other end with additional zip ties. The tubes remain taut because the zip ties cannot pass through the clamps. Once the tubes are secured in this manner, tape strips are unnecessary for the relatively low-speed parking measurements. This method significantly simplified and streamlined the many installations and removals that were needed throughout the year to effectively measure the utilization of the four parking lots.

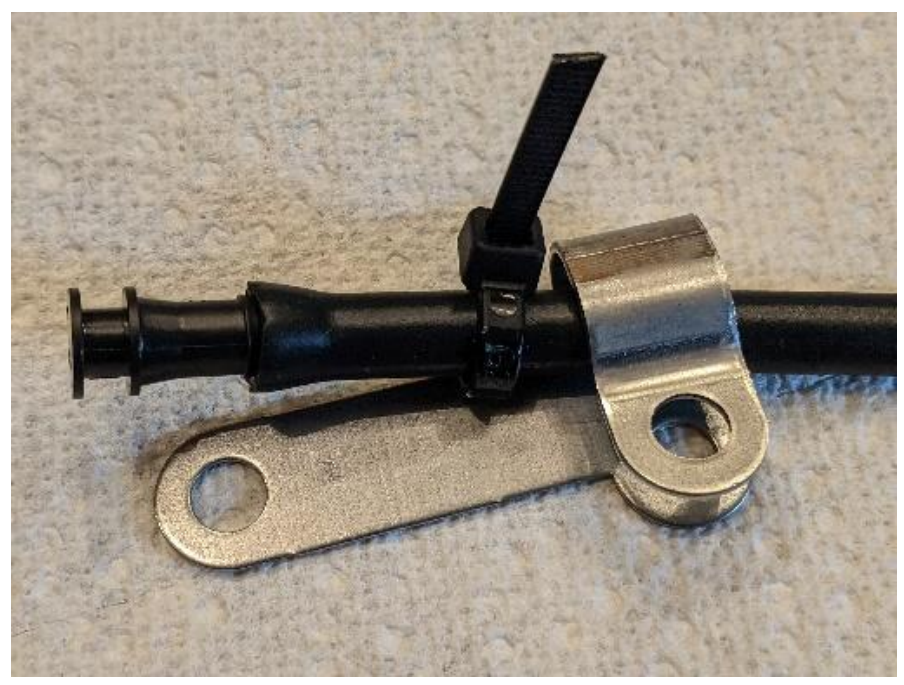

**Figure 7.** Use of plastic zip ties facilitated easy setup and removal of the tube system.

## Toqua

In addition to 170 spaces allocated for visitors, the Toqua parking lot includes a section reserved for POA trucks, trailers, and tractors. Visiting vehicles have been observed occupying some of these reserved spaces during overflow conditions. Since the tube system cannot distinguish which vehicles belong to which section of the lot, any maintenance vehicle that exits the lot early in the day and returns after the peak demand contributes to an underestimation of peak utilization for the visitor-allocated portion of the lot.

To address this issue, we prepared an alternate “backup” calculation that treated the maintenance section as part of the lot allocated for visitors. This estimate effectively increased the lot’s capacity by 5 spaces, with a margin of error of ±4 spaces. Due to the uncertainty of this value, the peak utilization observation error increased from approximately ±1.2% to ±3.1%. The expected demand for each night was increased by 2 vehicles—an estimate based on several observations of maintenance vehicles parking overnight. The observed peak demand of 148 exceeded the original count indicated in Table 1 by 5 vehicles. Analysis revealed that the peak demand error of -3 resulted from the presence of 5 vehicles in the maintenance section, rather than the expected 2, on the night following the peak demand observation.

Another challenge in measuring parking utilization at Toqua is that it is the only lot where golf carts constitute a non-negligible portion of traffic, estimated from timestamp data to reach as high as 4% on certain days. Golfers drive their carts through the lot to access the golf course but are not permitted to park in the lot. Therefore, it is necessary to exclude golf carts from parking demand counts.

Our early experiments revealed the difficulty of reliably classifying golf carts due to their wheelbase lengths overlapping with those of Class 1 vehicles (motorcycles, motor scooters, mopeds, motor-powered bicycles, and three-wheel motorcycles) and smaller Class 2 vehicles (passenger cars). Two-seater golf carts, which make up approximately 80% of carts in Tellico Village, have wheelbases ranging from 64 to 67 inches and were consistently classified as Class 1 vehicles in the experiments. Four- and six-seater carts have wheelbases ranging from 85 to 96 inches. STARnext classified some of these larger carts as Class 1 vehicles, while others remained unclassified, with the percentage varying depending on the D-Bounce setting.

Despite these classification issues, no modifications to the fine-tuning procedure described previously were necessary to filter out virtually all golf carts and achieve highly accurate counts. The optimal upper bound of the wheelbase range $w^*$ consistently exceeded 90 inches across the counts because the parameter optimization systematically discarded just enough valid detections when minimizing the cost function in Equation (8) to counterbalance invalid ones, whether they originated from cars or carts.

### Chota Recreation Center

The Chota Recreation Center has the smallest parking lot of the four facilities, with only 89 spaces. In the past, cars have occasionally been observed parking on the street outside the entrance. While this makes it challenging to measure the lot's demand at peak times, it is still useful to identify when and how often it exceeds capacity.

At the lot's entrance, distinct inbound and outbound lanes are separated by a raised median. A stop sign requires vehicles leaving the lot to come to a complete halt before turning onto the street. To accommodate these conditions, two counters were installed on opposite sides of the roadway in a staggered configuration, reducing the incidence of vehicles in the outbound lane stopping on the tubes. The counter on the inbound side used Lane 1 (A→B) to record vehicles entering the lot, while the counter on the outbound side used Lane 2 (A→B) to record vehicles exiting.

An additional challenge was the presence of an inner gravel section within the parking lot that is accessible almost immediately after vehicles enter the main area. This created a potential issue: cars crossing the tubes at slow speeds and steep angles—often including vehicles exiting through the entrance—generated a high number of unidentified pulses.

Another complication arose from the gradual drift in the exit counter's internal clock relative to that of the entrance counter. This timing discrepancy can affect demand estimates. The slower clock speed at the exit was likely caused by prolonged exposure to high mid-summer temperatures, whereas the entrance counter remained shaded and unaffected by direct sunlight.

Despite these challenges, the methods outlined in Sections 8-12 produced utilization estimates that were verified to be highly accurate, as shown in Table 1.

### Yacht Club

The Yacht Club has 179 parking spaces and two entrances: one providing direct access to the restaurant and the other leading to the marina. Aside from the need for a separate counter at each entrance, no special measures were required to determine the parking demand.

## 16. Conclusion

This paper has demonstrated a cost-effective approach to obtaining highly accurate parking utilization data without requiring extensive manual observation. Using a system of road tubes connected to portable traffic counters, the methodology provided a scalable and efficient means of gathering detailed vehicle count data.

The accuracy of the system depended on precise fine-tuning of parameters to minimize accumulated errors, followed by rigorous error correction. Automation of data exchange between the JAMAR STARnext analysis software and custom Python-based error calculation routines streamlined the parameter optimization process. Beyond accuracy, practical considerations were also addressed to streamline setup and removal of the tube system, reducing labor and minimizing disruptions to daily parking operations.

By leveraging a low-cost yet sophisticated methodology, this study at Tellico Village underscores the feasibility of achieving high-fidelity parking data without straining budgets. The insights gained from such an approach empower community planners to make data-driven decisions that enhance parking availability, improve traffic flow, and optimize facility usage, ensuring that as the community grows, its infrastructure remains well-equipped to support residents and visitors.

## Appendix: Buffer Calculations

For each day and rolling interval, we predicted the expected excess demand at buildout. An empirical cumulative distribution function (CDF) was then used to estimate the buffers required to meet various daily and rolling coverage targets. Figure A.1 indicates that, at buildout, a buffer of 39 cars or spaces will be needed to cover 95% of days at the Yacht Club, whereas a buffer of 41 will cover 99.5% of 5-minute intervals.

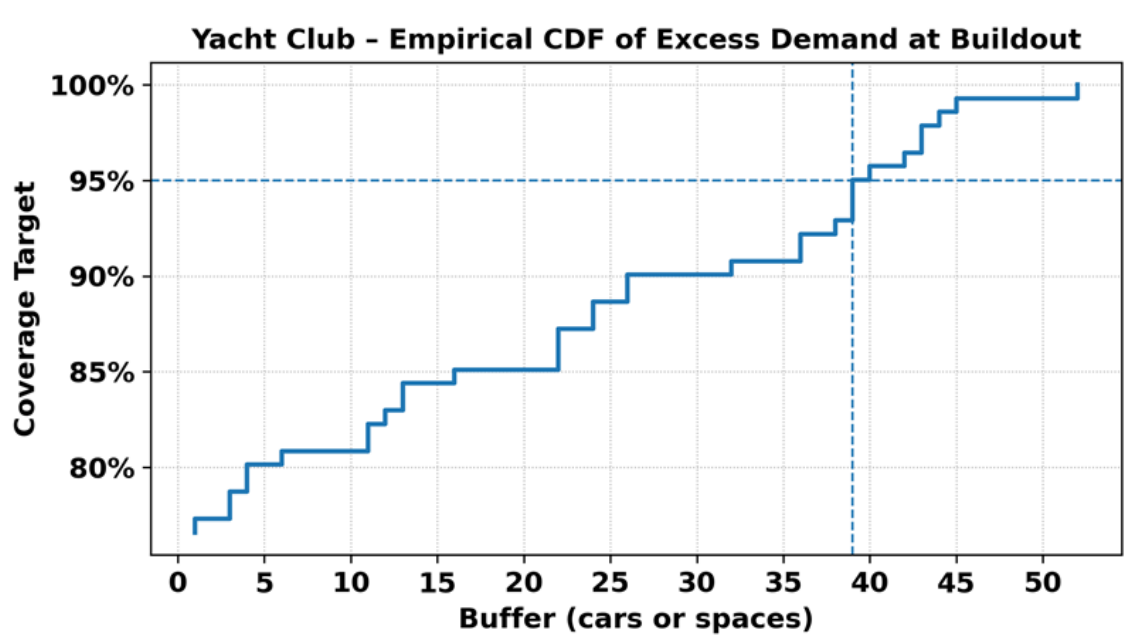

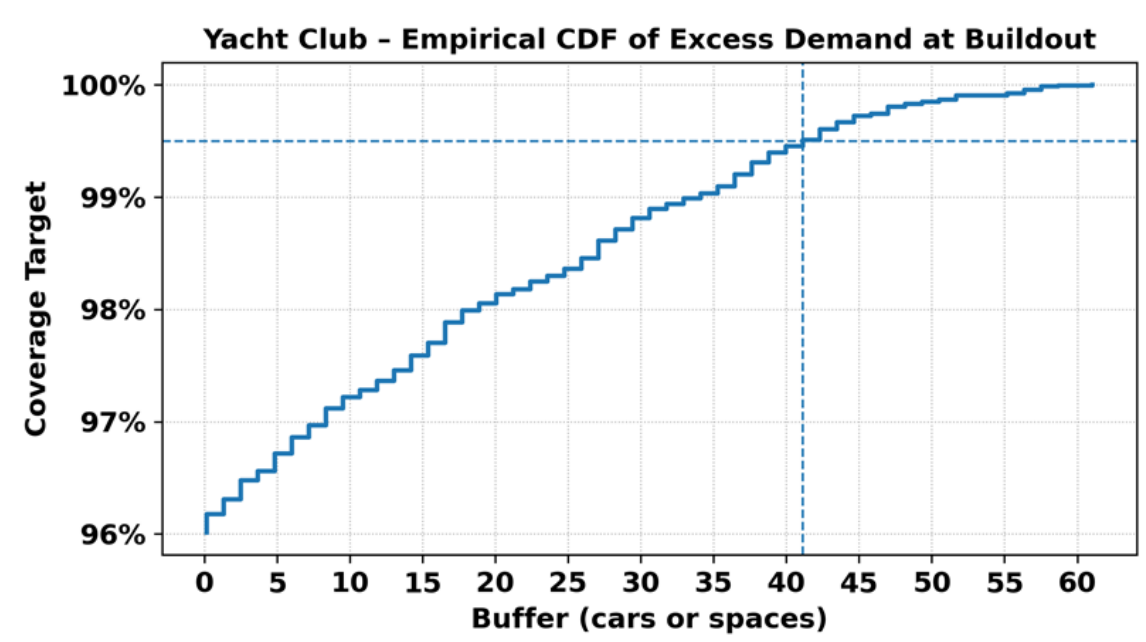

**Figure A.1.** Empirical cumulative distribution functions showing the buffer requirements needed to meet daily coverage targets (left) and rolling 5-minute coverage targets (right).

The CDF indicates the probability that excess demand is less than or equal to any given value. It is constructed from each interval's contribution of $1/\mathrm{N}$ to the cumulative share, where $\mathrm{N}$ is the total number of intervals. If $\mathrm{p}$ represents the proportion of intervals for which peak utilization is less than or equal to the current buildout percentage, then the remaining intervals correspond to those with positive excess demand at buildout. The cumulative share for the $\mathrm{k}$th interval of this latter set is therefore:

$$(\mathrm{A}.1) \quad \mathrm{p} + \frac{\mathrm{k}}{N}$$

We define the coverage target $q$ as the minimal acceptable proportion of intervals for which excess demand is zero at full buildout. The $k$th interval reaches this coverage target if:

$$(A.2) \quad k \geq N(q - p)$$

The excess demand estimates at buildout are ordered from smallest to largest. The first interval that satisfies Equation (A.2) indexes the excess demand value $D_k$, which represents the minimum buffer required to meet the specified coverage target.

Upper estimates were also reported to help facility managers plan for severe demand spikes or unanticipated population growth. The range of buffer values was obtained from grid sweeps that varied both the assumed current buildout percentage and the percentage of days currently exceeding capacity.

## Acknowledgements

I sincerely appreciate the contributions of Tellico Village LRPAC volunteers Domenick Andreana, Bruce Palansky, and James Stutz. I would also like to thank Igor Marques and Jonathan Milanese of JAMAR Technologies for their valuable support of the STARnext software solution.